\documentstyle[12pt,graphicx]{article}

\begin{document}

\title{Images and Spectra From the Interior 
  of a Relativistic Fireball}
\author{Jonathan Granot, Tsvi Piran and  Re'em Sari}
\maketitle
%\date{}
%\affil{Racah Institute, Hebrew University, Jerusalem 91904, Israel }
%\authoremail{
%jgranot@nikki.fiz.huji.ac.il
%tsvi@nikki.fiz.huji.ac.il
%sari@nikki.fiz.huji.ac.il}

\begin{abstract}
  The detection of an afterglow, following a $\gamma$-ray burst (GRB),
  can be described reasonably well by synchrotron emission from a
  relativistic spherical expanding blast wave, driven by an expanding
  fireball. We perform detailed calculations considering the emission
  from the whole region behind the shock front. We use the Blandford
  McKee self similar adiabatic solution to describe the fluid behind
  the shock. Using this detailed model, we derive expressions for the
  peak flux, and the peak frequency at a given observed time. These
  expressions provide important numerical corrections to previous,
  more simplified models. We calculate the observed light curve and
  spectra for several magnetic field models. We show that both the
  light curve and the spectra are flat near the peak. This rules out
  the interpretation of the optical peak of GRB970508 as the peak of
  the light curve, predicted by the existing fireball models. We
  calculate the observed image of a GRB afterglow. The observed image
  is bright near the edge and dimmer at the center, thus creating a
  ring. The contrast between the edge and the center is larger at high
  frequencies and the width of the ring is smaller.
\end{abstract}

%\keywords{gamma rays: bursts; hydrodynamics; relativity; shock waves}

\section{Introduction}
The detection of delayed X-ray, optical and radio emission,
``afterglow'', following a GRB is reasonably described by emission
from a spherical relativistic shell, decelerating upon collision with
an ambient medium (Waxman 1997a, M\'esz\'aros \& Rees 1997, Katz \&
Piran 1997, Sari, Piran \& Narayan 1998). A relativistic blast wave is
formed and expands through the surrounding medium, heating the matter
in it's wake. The observed afterglow is believed to be due to
synchrotron emission of relativistic electrons from the heated matter.
The surrounding medium will be referred to as interstellar medium
(ISM), though this may not necessarily be the case.

At any given time, a detector receives photons which were emitted at
different times in the observer frame, at different distances behind
the shock front and at different angles from the line of sight (LOS)
to the center of the GRB. The properties of the matter are different
at each of these points, and so are the emissivity and the frequency
of the emitted radiation. Early calculations have considered emission
from a single representative point (M\'esz\'aros \& Rees 1997, Waxman
1997a, Sari, Piran \& Narayan 1998). Later works included more
detailed calculations.  Synchrotron emission was considered from the
shock front (Sari 1998, Panaitescu \& M\'esz\'aros 1998), and
monochromatic emission was considered from a uniform shell (Waxman
1997b).

We consider an adiabatic hydrodynamical evolution, and use the
adiabatic self similar solution found by Blandford and McKee
(1976) for a highly relativistic blast wave expanding into a
uniform cold medium, which we will refer to as the BM solution. We
neglect scattering, self absorption and electron cooling.  Self
absorption becomes important at frequencies much smaller than the peak
frequency, and for slow cooling, electron cooling becomes important at
frequencies much higher than the peak frequency, so this should yield
a good approximation for the observed flux around the peak. An
analysis of the spectrum over a wider range of frequencies was made by
Sari, Piran and Narayan (1998).

In section 2 we derive the basic formula for the observed flux from a
system moving relativistically. In section 3 we consider synchrotron
emission from a power law electron distribution, and calculate the
observed light curve and spectra. We show that both the light curve
and the spectra are flat near the peak. This causes difficulty in
explaining the shape of the optical peak of GRB970508 (Sokolov et. al.
1997). We obtain expressions for the peak frequency at a given
observed time, and for the peak flux. In section 4 we consider three
alternative magnetic field models. We show that the light curve and
the spectra remain flat near the peak in all the cases we considered.

In section 5 we calculate the observed light curve and spectra from a
locally monochromatic emission. We consider a uniform shell
approximation, which was calculated by Waxman (1997c), and compare the
light curve and spectra to those obtained for the BM solution. We show
that a uniform shell approximation yields results which are
considerably different from those obtained for a more realistic
hydrodynamics, and very different from those obtained for a realistic
emission together with a realistic hydrodynamics.

In section 6 we calculate the surface brightness, thus obtaining the
observed image of a GRB afterglow. As indicated in previous works
(Waxman 1997b), Sari 1998, Panaitescu \& M\'esz\'aros 1998), we obtain
from detailed calculations, that the image appears brighter near the
edge and dimmer near the center, creating a ring near the outer edge.
At a given observed time, the contrast between the edge and the center
of the image is larger and the width of the ring is smaller at high
frequencies, while at low frequencies the contrast is smaller and the
width of the ring is larger.

\section{The Formalism}
\begin{figure}
\centering
\noindent
\includegraphics[width=13cm]{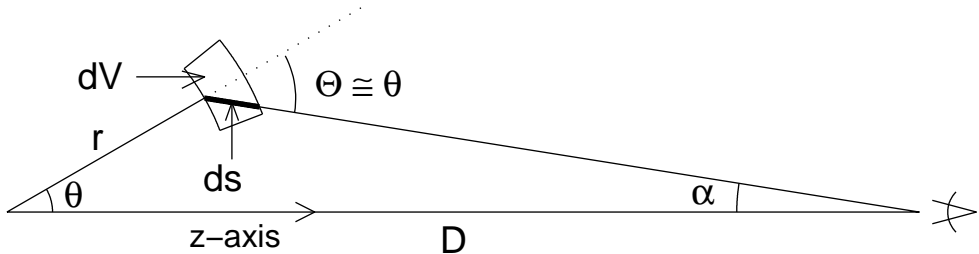}
\caption{\label{Fig1}
  The contribution of a volume element $dV$ to the flux observed by a
  distant observer is $dF_{\nu} \approx I_\nu d\Omega$, where
  $I_{\nu}=j_{\nu}ds$. Since the observer is far away, the direction
  of emission in the observer frame is almost parallel to the z-axis.}
\end{figure}

We consider a system that is moving relativistically while emitting
radiation.  We obtain a formula for the flux that is measured by a
distant detector (i.e. $D \gg L$ where D is the distance to the
detector, and L is the size of the area emitting radiation).  We use a
spherical coordinate system and place the origin within the emitting
region (i.e. at the source), while the z-axis points towards the
detector (see Figure \ref{Fig1}).  The detector is at rest in this
frame, and so is the ambient ISM. We refer to this frame as the
observer frame.  Consider a small volume element $dV=r^2drd\mu d\phi$
(where $\mu \equiv \cos \theta$) and let $j'_{\nu'}$ be the energy per
unit time per unit volume per unit frequency per unit solid angle,
emitted by the matter within this volume in it's local frame (note
that generally $j_\nu$ depends on the direction $\Omega$ as well as on
the frequency, place and time). We denote quantities measured in the
local rest frame of the matter with a prime, while quantities without
a prime are measured in the observer frame. Note that $j_\nu /\nu^2$
is Lorentz invariant (Rybici \& Lightman 1979) and $\nu'=\nu
\gamma(1-\beta \mu_v)$, where $\gamma$ and $\beta c$ are the Lorentz
factor and the velocity of the matter emitting the radiation,
respectively, and $\mu_v$ is the cosine of the angle between the
direction of the velocity of the matter and the direction to the
detector, in the observer frame). The contribution to $I_{\nu}$ from
this volume element is given by:
\begin{equation}
\label{dI nu}
dI_\nu \cong j_\nu {dr \over \mu}={j'_{\nu'} \over \gamma^2(1-\beta
\mu_v)^2}{dr \over \mu}
\end{equation}
(see Figure \ref{Fig1}). The contribution to the flux at the detector
is $dF_\nu \cong I_\nu d\Omega $ where $\Omega$ is the solid angle
seen from the detector, and $I_\nu$ includes all the contributions
from different volume elements along the trajectory arriving at the
detector from the direction $\Omega$ simultaneously and at the time
for which $F_\nu$ is calculated. Consider a photon emitted at time $t$ and
place $\vec {\bf r}$ in the observer frame. It will reach the detector
at a time $T$ given by;
\begin{equation}
\label{time}
T=t -{r \mu \over c} \ ,
\end{equation}
where $T=0$ was chosen as the time of arrival at the detector of a
photon emitted at the origin at $t=0$. Using $\alpha \cong
r\sqrt{1-\mu^2} / D $, we obtain:
\begin{equation}
\label{flux j}
F(\nu , T)={1 \over D^2}\int^{2\pi}_0d\phi \int^1_{-1}d\mu
\int^{\infty}_0r^2dr{{j'(\Omega'_d , \nu \gamma(1-
\beta\mu_v) , {\vec {\bf r}} , T+{r \mu \over c})} \over
\gamma^2(1- \beta \mu_v)^2}\ ,
\end{equation}
where $j'=j'(\Omega' , \nu' , {\vec {\bf r}} , t )$, and later $P'=P'(
\nu' , {\vec {\bf r}} , t )$. $j'$ is taken in the direction
$\Omega'_d$ at which a photon should be emitted in order to reach the
detector, and $\gamma, \beta, \mu _v$ should be taken at the time $t$
implied by equation \ref{time}.

For a spherical expanding system, which emits isotropically in it's
local rest frame, one obtains $\mu_v=\mu$ and $j'_\nu=P'_\nu / 4 \pi$,
so that:
\begin{equation}
\label{preFAF}
F(\nu , T)={1 \over 2D^2}\int^1_{-1}d\mu
\int^{\infty}_0r^2dr{{P'(\nu \gamma(1- \beta\mu) , r ,
T+{r \mu \over c})} \over \gamma^2(1- \beta \mu)^2} \ .
\end{equation}
Note that because of relativistic effects, a jet with an opening angle
$\theta>1/\gamma$ around the LOS can be considered locally as
spherical (Piran 1994).

In order to learn whether the radial integration is important, we
calculate the observed flux from emission only along the LOS. We do
this by considering a situation in which at each point the photons are
emitted only radially: $j'_{\nu'}=P'_{\nu'}\delta(\Omega'-
\Omega'(\hat r))$. Note that the correct limit is obtained when the
delta function in the direction of the emission is taken in the local
frame. Since $d\Omega=\gamma^2(1-\beta \mu)^2d\Omega'$ (Rybici \&
Lightman 1979) we obtain that:
\begin{equation}
\delta(\Omega'- \Omega'(\hat r))=\gamma^2 (1-\beta \mu)^2 \delta(\Omega- \Omega(\hat r)) \ . 
\end{equation}
Substituting this $j'$ in equation \ref{flux j} we obtain:
\begin{equation}
\label{preLOS}
F(\nu ,T)={1 \over D^2}\int^{\infty}_0r^2dr{P'({\nu \over
\gamma(1+\beta)} , r , T+{r \over c})} \ .
\end{equation}

\begin{figure}
\centering
\noindent
\includegraphics[width=13cm]{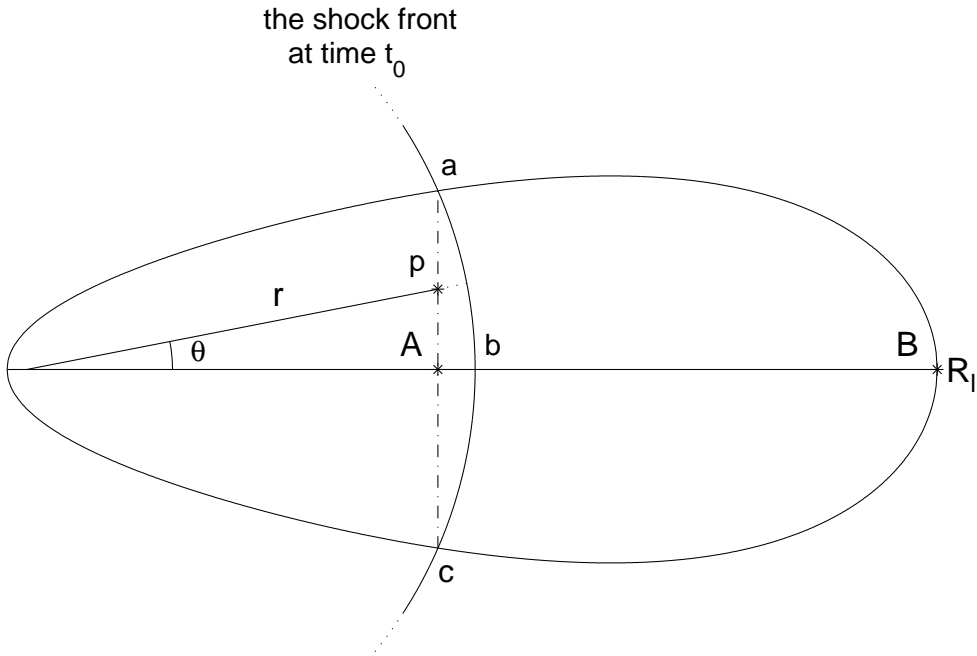}
\caption{\label{Fig2}
  We see here the egg-shaped region from which photons reach an
  observer at a given time $T$. In all the figures of the ``egg'' the
  observer is located far to the right, and the symmetry axis is the
  LOS to the center of the GRB. In order to reach the observer
  simultaneously, photons emitted at different locations should be
  emitted at different times in the observer frame, according to
  equation \ref{time}.  Photons that are emitted simultaneously in the
  observer frame along the dashed-dotted line $ac$, reach the observer
  simultaneously. Therefore $ac$ represents an equal $y$ contour line.
  The location of the shock front at this time of emission is
  indicated by the solid line $abc$, and it's radius $R$ appears in
  the definitions of $y,$ and $\chi$. A photon emitted at point A, at
  a distance of $d=r/\gamma^2$ behind the shock front and at a time
  $t_0$ (both in the observer frame) caches up with the shock front at
  point B, at a later time $t_1$ in the observer frame.}
\end{figure}

Equation \ref{preFAF} is quite general, and includes integration over
all space. In the case of GRB afterglow, radiation is emitted only
from the region behind the shock front. The spacial integration should
therefore be taken over a finite volume, confined by the surface of
the shock front. We would therefore like to obtain an explicit
expression for the radius of the shock $R$ as a function of $\mu
\equiv \cos \theta$ for a given arrival time $T$. In the case of a
shell moving with a constant velocity $\beta c$, one obtains from
equation \ref{time}:
\begin{equation}
\label{ellipse}
R={\beta c T \over 1 - \beta \mu}\ .
\end{equation}
If one considers a constant arrival time $T$, this equation describes
an ellipsoid, which confines the volume constituting the locus of
points from which photons reach the detector simultaneously (Rees
1966). In GRBs, most of the matter is concentrated in a thin shell
which decelerates upon collision with the ambient medium. When the
deceleration of the shell is accounted for, the ellipsoid is
distorted. The details of this distortion depend on the evolution of
the shock radius $R(t)$ (Sari 1998, Panaitescu \& M\'esz\'aros 1998).
In this paper we consider an adiabatic ultra-relativistic hydrodynamic
solution, which implies $\Gamma \propto R^{-3/2}$, where $\Gamma$ is
the Lorentz factor of the shock. For this case, equation \ref{time}
yields:
\begin{equation}
\label{egg}
R={c T \over 1 - \mu + 1/(8 \Gamma^2)}\ .
\end{equation}
The shape of the volume constituting the locus of points from which
photons reach the detector simultaneously resembles an elongated egg
(see Figure \ref{Fig2}), and will be simply referred to as the
``egg''. The side facing the observer (the right side in the Figures)
will be referred to as the front of the ``egg'', and the side closer to 
the origin (the left side in the figures) will be referred to as the
back of the ``egg''.

From hydrodynamic considerations, we expect the typical size of the
region emitting radiation behind the shock, to scale as: $\Delta
\propto R/\Gamma^2$ in the observer frame. Despite this fact, it is
still important to consider the emission from the whole volume of the
``egg'', whose limits are given by equation \ref{egg}. To illustrate
this we give a simple example. Consider a photon emitted on the LOS at
a distance $d=R/\gamma^2$ behind the shock (point A in Figure
\ref{Fig2}) at a time $t_0$ in the observer frame, where
$\gamma=\Gamma/\sqrt 2$ is the Lorentz factor of the matter just
behind the shock. This Photon will catch up with the shock front at a
later time $t_1$ in the observer frame (at point B in Figure
\ref{Fig2}), and will arrive at the detector together with a photon
emitted at that point at $t_1$. From equation \ref{egg} we find that
on the LOS: $R=16\gamma^2 c T$, and so we obtain that
$R(t_1)/R(t_0)=17^{1/4} \cong 2$. This shows that the emission comes
from a substantial part of the volume of the ``egg'', and not just
from a thin layer near it's surface. This is illustrated by the shaded
region in figures \ref{Fig5} and \ref{Fig6}, which corresponds to a
shell of width $\Delta=R/4\gamma^2$ in the observer frame.

\begin{figure}
\centering
\noindent
\includegraphics[width=9cm]{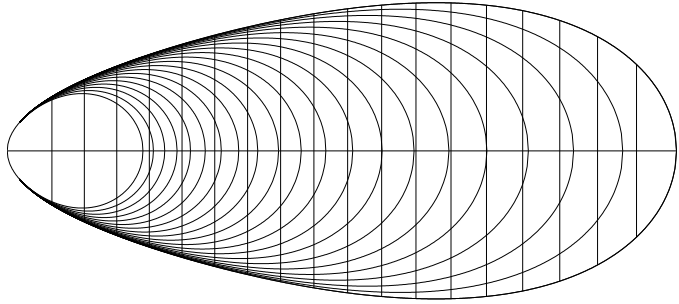}
\caption{\label{Fig3}
  Equal-$y$ contour lines ($y=0.1,0.15,...,0.95$) - vertical lines,
  and equal-$\chi$ contour lines ($Log_{10}(\chi)=0,0.15,0.3,...,3$) -
  curved lines. The horizontal line is the LOS to the observer, which
  is located far to the right.}
\end{figure}

At a given observed time $T$, the emission should be considered from
the volume of the ``egg'' whose surface is described by equation
\ref{egg}. Taking this into account, it is simpler to calculate the
flux at a given observed time $T$, using new variables $y,\chi$ that
depend on $T$ (see Figures \ref{Fig2} and \ref{Fig3}):
\begin{equation}
\label{def y chi}
y \equiv {R \over R_l}     \ \ \ \ \ \ \ ,\ \ \ \ \ \ \ 
\chi \equiv 1+16 \gamma_f^2\left({R-r \over R}\right) \ ,
\end{equation}
where $R=R(t)$ is the radius of the shock front, $\gamma_f$ is the
Lorentz factor of the matter just behind the shock and $R_l$ is the
radius of the point on the shock front, on the LOS from which a photon
reaches the detector at a time $T$ (see Figure \ref{Fig2}). Since we
expect the typical size of the emitting region behind the shock, to
scale as: $\Delta \propto R/\gamma^2$, the choice of $\chi$ is natural
to this problem. The exact form of $\chi$ was chosen to suit the BM
solution, discussed below. Using equation \ref{time} we can express $r,\mu$
in terms of $y,\chi$:
\begin{equation}
\label{r,mu to y,chi}
r \cong R_ly    \ \ \ \ \ \ \ \ ,\ \ \ \ \ \ \ \ 
\mu \cong 1-{1-\chi y^4 \over 16\gamma_l^2y}\ ,
\end{equation}
where $\gamma_l$ is the Lorentz factor of the matter just behind the
shock, on the LOS.

We would like to express equation \ref{preFAF} in terms of $y,\chi$.
This will enable us to calculate the flux for the BM solution. This
solution describes an adiabatic highly relativistic blast wave
expanding into an ambient uniform and cold medium (Blandford \& McKee
1976). In terms of $y$ and $\chi$ the BM solution is given by:
\begin{equation}
\label{BM}
n'= 4 \gamma_f n_1 \chi^{-5/4} \ \ \ ,\ \ \ 
\gamma=\gamma_f \chi^{-1/2} \ \ \ ,\ \ \ 
e'=4 n_1 m_p c^2 \gamma_f^2 \chi^{-17/12} \ ,
\end{equation}
where $n'$ and $e'$ are the number density and the energy density in
the local frame, respectively, $\gamma$ is the Lorentz factor of the
bulk motion of the matter behind the shock, $m_p$ is the mass of a
proton, $n_1$ is the number density of the unshocked ambient ISM in
it's local rest frame and $\gamma_f=\gamma_ly^{-3/2}$. For the BM
solution, one obtains (Sari 1997):
\begin{equation}
\label{BM2}
\gamma_l \cong 3.65E_{52}^{1/8}n_1^{-1/8}T_{days}^{-3/8} \quad, \quad
R_l \cong 5.53\times 10^{17}E_{52}^{1/4}n_1^{-1/4}T_{days}^{1/4}\rm{cm} \ ,
\end{equation}
where $E_{52}$ is the total energy of the shell in units of
$10^{52}\rm {erg}$, $n_1$ is the number density of the ISM in units of
$\rm cm^{-3}$ and $T_{days}$ is the observed time in days.

For any spherically symmetric self similar solution we can define
$g(\chi)$ by: $\gamma^2 \equiv \gamma_f^2g(\chi)$, where $g(\chi)$
describes how $\gamma$ varies with the radial profile.  For the BM
solution $g(\chi)=\chi^{-1}$, and for a uniform shell $g(\chi)=1$.
Using the definitions above, we obtain from equation \ref{preFAF}
after the change of variables:
\begin{equation}
\label{FAF}
F(\nu , T)={8 R_l^3 \over D^2}\int^{\infty}_1d\chi
\int^{\chi^{-1/4}}_0 dy {y^{10}P'(T ,\nu \gamma(1- \beta\mu), y
  , \chi) \over {\left[1+y^4\left( 8 g(\chi)^{-1}-\chi \right)
    \right]^2 g(\chi)}} \ .
\end{equation}
This formula for the flux takes into consideration the contribution
from the whole volume behind the shock front, and will be referred to
as the general formula.

We would like to single out the angular integration and the radial
integration, in order to find out the different effects each
integration has on the observed flux.

In order to single out the effect of the angular integration we
consider a thin shell of thickness $\Delta$ in the observer frame and
take the limit $\Delta \to 0$. Because of kinematical spreading we
expect $\Delta$ to scale as: $\Delta\propto R/\gamma^2$. According to
the definition of $\chi$, such a shell corresponds to a constant
interval in $\chi$ , i.e. the shell lies within the interval $\chi \in
[1,\chi_{max}]$ for some $\chi_{max}$. The limit $\Delta \to 0$
corresponds to taking a delta function in $\chi$ : $\delta(\chi - 1)$.

In order to single out the effect of the radial integration, we
changed variables in equation \ref{preLOS} from $r$ to $\chi$. For the
BM solution we obtain:
\begin{equation}
\label{LOS}
F(\nu , T)={R_l^3 \over 4 D^2}\int^{\infty}_1d\chi \chi^{-7/4} 
P'(T , {\nu \over \gamma(1+\beta)}, \chi ) \ .
\end{equation}

\section{Synchrotron Emission}
\label{ni3}
According to the fireball model, a highly relativistic shell moves
outward and is decelerated upon collision with the ambient ISM. This
creates a relativistic blast wave that expands through the ISM and
heats up the matter that passes through it. The relativistic electrons
of the heated material emit synchrotron radiation in the presence of a 
magnetic field.

We now consider the synchrotron emission at a certain point, in which
the values of $n',\gamma$ and $e'$ are given by the hydrodynamic
solution. In order to estimate the local emissivity, we assume that
the energy of the electrons and of the magnetic field at each point
are a fixed fraction of the total internal energy at that point:
$e'_{el}=\epsilon_e e'$ , $e'_B=\epsilon_B e'$. We assume that the
shock produces a power law electron distribution:
$N(\gamma_e)=K{\gamma_e}^{-p}$ for $\gamma_e \ge \gamma_{min}$ (for the
energy of the electrons to remain finite we must have $p>2$). In the
Figures for which a definite numerical value of $p$ is needed, we use
$p=2.5$. The constants $K$ and $\gamma_{min}$ in the electron
distribution can be calculated from the number density and energy
density:
\begin{equation}
\label{el dis}
\gamma_{min}=\left({p-2\over p-1}\right){\epsilon_e e' \over n'm_ec^2} 
\ \ \ \ \ \ ,\ \ \ \ \ \ 
K=(p-1) n' \gamma_{min}^{p-1} \ .
\end{equation}

Assuming an isotropic velocity distribution, the total emitted power
of a single average electron (i.e. with $\gamma_e=\langle \gamma_e
\rangle$) is given by:
\begin{equation}
\label{P el}
P'_{e,av}={4 \over 3} \sigma_T c \beta^2 \langle \gamma_e\rangle^2 e'_B
\quad 
, \quad \langle \gamma_e \rangle \equiv {\epsilon_e e' \over n' m_e c^2} 
\end{equation}
(Rybici \& Lightman 1979), where $\sigma_T$ is the Thomson cross
section and $e'_B=B^2 / 8 \pi$, where $B$ is the magnetic field (in
the local frame). Although we refer to the magnetic field in the local
frame, throughout the paper, we make an exception and write it without
a prime.

The synchrotron emission function (power per unit frequency) of a
single electron is characterized by $P_{\nu,e} \propto \nu^{1/3}$ for
frequencies much smaller than the electrons synchrotron frequency, and
it drops exponentially at large frequencies. The typical synchrotron
frequency, averaged over an isotropic distribution of electron
velocities, is given by:
\begin{equation}
\label{typical syn freq}
\nu'_{syn}(\gamma_e)={3 \gamma_e^2 q_e B \over 16 m_e c}\ ,
\end{equation}
where $m_e$ and $q_e$ are the mass and the electric charge of the
electron, respectively. We approximate the emission of a single
electron as $P_{\nu,e} \propto \nu^{1/3}$ up to the electrons typical
synchrotron frequency, where we place a cutoff in the emitted power.
We normalized the emission function so that the total power emitted by
a single electron equals that of an exact synchrotron emission.

\begin{figure}
\centering
\noindent
\includegraphics[width=9cm]{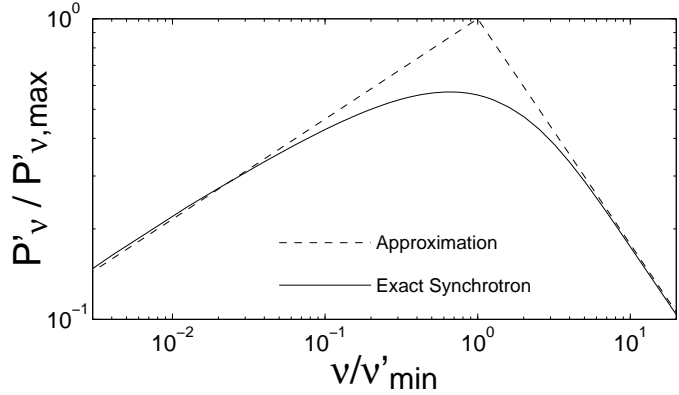}
\caption{\label{Fig17}
  The local emissivity from a power law distribution of electrons,
  emitting synchrotron radiation. The solid curve represents exact
  synchrotron emission with an isotropic electron velocity
  distribution, while the dashed curve represents the approximation we
  used for the local emissivity. The differences in the local
  emissivity tend to smear out when integration is performed over the
  whole volume behind the shock front.}
\end{figure}

Under these assumptions, we obtain after integration over the power
law electron distribution, that the spectral power per unit
volume (in the local frame) at any given point is:
\begin{equation}
\label{emission function} 
P'_{\nu}=\left\{\matrix{ P'_{\nu,max}\left( {\nu \over
\nu'_{min} }\right)^{1/3} & \nu<\nu'_{min} \cr 
P'_{\nu,max}\left( {\nu \over 
\nu'_{min} }\right)^{-{p-1 \over 2}} & 
\nu>\nu'_{min}} \right. \quad ,
\end{equation}
where $\nu'_{min}=\nu'_{syn}(\gamma_{min})$ is the synchrotron
frequency of an electron with the minimal Lorenz factor at that point.
Since the emitted power at each point peaks at $\nu'_{min}$, this
frequency can be looked upon as the typical emitted frequency around
which the emitted power is concentrated. Although this emission
function was obtained by approximating the spectral emission of each
electron as having the shape of the low frequency tail ($\propto
\nu^{1/3}$), the spectral power for the whole electron distribution
resembles that obtained for an exact synchrotron emission and an
isotropic electron velocity distribution. The solid curve in Figure
\ref{Fig17} represents the local emissivity from an exact synchrotron
emission of an isotropic distribution of electrons (Rybici \& Lightman
1979, Wijers \& Galama 1998), while the dashed curve represents
equation \ref{emission function} with:
\begin{equation}
\label{0.88}
P'_{\nu,max}=0.88 \times {4(p-1) \over (3p-1)}
{n' P'_{e,av} \over \nu'_{syn}(\langle \gamma_e \rangle)}, 
\end{equation}
where a factor of $0.88$ was added to improve the fit to the exact
synchrotron emission.  In our calculations we use the local emission
represented by the dashed line in Figure \ref{Fig17}. This local
emission differs from that of an exact synchrotron emission near the
peak, by up to $\sim 45\%$, and is only slightly different above or
below the peak. Note that differences in the local emissivity tend to
get smeared out, when the contribution to the observed flux is
integrated over the whole volume behind the shock front. We expect
that considering exact synchrotron emission should somewhat lower the
peak flux and the peak frequency, and make the light curve and spectra
more rounded and flat near the peak. We evaluate that the peak flux
would be lower by about $30\%$. Since there are only slight
differences in high and low frequencies, there should hardly be any
effect on the value of the point where the extrapolations of the power
laws at high and low frequencies meet (see Table 1).

In this section we use the BM hydrodynamical solution. The combination
of the realistic emission function described above, and a realistic
hydrodynamical solution should yield a reasonable approximation for
the observed flux.

The results are presented as a function of the dimensionless
similarity variable $\phi \equiv \nu/\nu_{T}$, where $\nu_{T}$ is
defined as the observed synchrotron frequency of an electron with
$\gamma_e=\gamma_{min}$ just behind the shock on the LOS:
\begin{equation}
\label{nuT}
\nu_{T}\equiv \nu_{min}(y=\chi=1)=1.54 \times 10^{15} \sqrt {1+z} 
\left({p-2 \over p-1}\right)^2 \epsilon_B^{1/2} \epsilon_e^2
E_{52}^{1/2} T_{days}^{-3/2}\rm Hz
\end{equation}
where $z$ is the cosmological red shift of the GRB. Our results can
therefore be looked upon, with the proper scaling of the logarithmic
x-axis, either as the spectra at a given observed time or as the light
curve at a fixed observed frequency.

Similarly, we express the observed flux in terms of a ``standard''
flux $F_0$ defined by:
\begin{equation}
\label{def F_0}
F_0 \equiv {N(R_l) P_{\nu,peak} \over 4 \pi D^2} \quad , \quad  
P_{\nu,peak} \equiv 0.88 {P_{e,av}(R_l)\over \nu_{syn}(\gamma_l)} = 
0.88 {P'_{e,av}(R_l)\gamma_l \over \nu'_{syn}(\gamma_l)} \ ,
\end{equation}
where $P'_{e,av}(R_l)$ and $ \nu'_{syn}(\gamma_l)$ are the total power
and synchrotron frequency of an average electron at $y=\chi=1$,
respectively, $P_{\nu,peak}$ is an estimate of the peak spectral power
of an average electron, and $N(R_l)\equiv 4 \pi R_l^3 n_1 /3$ is the
total number of electrons behind the shock at the time $t$ in the
observer frame for which $R(t)=R_l$. Allowing for cosmological
corrections, $F_0$ is given by:
\begin{equation}
\label{F_0}
F_0=8.24\times 10^4 (1+z) \epsilon_B^{1/2} E_{52} n_1^{1/2}
 \left({d_L \over 10^{28}\rm{cm}}\right)^{-2} \rm{\mu J} \ ,
\end{equation}
where $d_L$ is the luminosity distance of the GRB.

\begin{figure}
\centering
\noindent
\includegraphics[width=10cm]{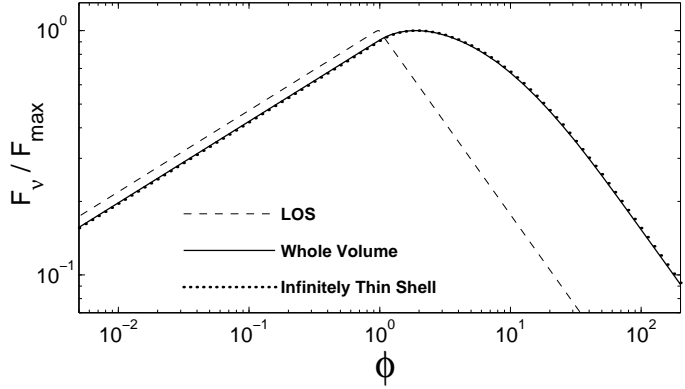}
\caption{\label{Fig8}
  The observed flux from synchrotron emission of a power law electron
  distribution for the BM solution. The different curves stand for
  emission only along the LOS, emission from an infinitely thin shell,
  and emission from the whole volume behind the shock. The emission
  from the whole volume is very similar to that obtained for an
  infinitely thin shell, due to a coincidence that rises from our
  choice for the magnetic field model (see section \ref{difB}).}
\end{figure} 

The results for the observed flux are presented in Figure \ref{Fig8}.
The flux arriving from the LOS (the dashed curve) peaks at
$\phi_{peak} \cong 0.98$, and is only slightly rounded near the peak.
The flux arriving from an infinitely thin shell (the dotted curve)
peaks at $\phi_{peak} \cong 1.92$ and is quite rounded and flat near
the peak. The flux arriving from the whole volume behind the shock
front (the solid curve) peaks at $\phi_{peak} \cong 1.88$ and is flat
and rounded near the peak, quite resembling the flux from an
infinitely thin shell.   

Our best prediction for the flux measured by a distant detector is
obtained when we assume a realistic hydrodynamical solution (the BM
solution), a realistic radiation emission and use the general formula
(see the solid curve in Figure \ref{Fig8}). We would now like to
examine it more closely. The curve looks quite flat near the peak, and
we attempt to demonstrate this feature in a quantitative manner. If
one compares the peak flux obtained by extrapolation of the power
laws, obtained at low and high frequencies, to the ``actual'' peak
flux, one obtains that it is larger by a factor of $1.53$:
$F_{extr}=1.53F_{\nu,max}$. In order to further estimate the flatness
of the curve near the peak, we define $\phi_+$ and $\phi_-$ by
$F_{\nu}(\phi_+)=F_{\nu}(\phi_-)=0.5F_{\nu,max}$, where
$\phi_+>\phi_{peak}>\phi_-$.  We obtain that $\phi_- \approx 0.1
\phi_{peak}$ and $\phi_+ \approx 10 \phi_{peak}$, so that
$\phi_+/\phi_- \sim 100$. In a similar manner, we define $T_+$ and
$T_-$ for a given observed frequency, by
$F_{\nu}(T_+)=F_{\nu}(T_-)=0.5F_{\nu,max}$, where $T_+>T_{peak}>T_-$.
One obtains: $T_+/T_-=(\phi_+/\phi_-)^{2/3} \sim 20$ (note that $T_+$
and $T_-$ are frequency independent).

The frequency at which the observed flux peaks at a given observed
time $T$ is given by:
\begin{equation}
\label{ni peak}
\nu_{peak}=\phi_{peak}\nu_T \cong 2.9\times10^{15}\sqrt{1+z}\left({p-2 \over
    p-1}\right)^2 
\epsilon_B^{1/2} \epsilon_e^2 E_{52}^{1/2} T_{days}^{-3/2}\rm{Hz} \ .
\end{equation}
The main dependence on $p$ lies in the factor $(p-2)^2/(p-1)^2$, but
there is also a weak dependence of $\phi_{peak}$ on $p$. The value
given here for $\phi_{peak}$ is for $p=2.5$, and it decreases with
increasing $p$ (it is higher by $24\%$ for $p=2$, and lower by $12\%$
for $p=3$).  The maximal flux is given by:
\begin{equation}
\label{Fmax}
F_{\nu,max} \cong 1.7\times10^4(1+z) \epsilon_B^{1/2}E_{52}n_1^{1/2} 
 \left({d_L \over 10^{28}\rm{cm}}\right)^{-2} \rm{\mu J} \ .
\end{equation}
Here there is a much weaker dependence on $p$. The value stated
here is for $p=2.5$, and it is smaller by $11\%$ for $p=2$ and larger
by $6\%$ for $p=3$.

\begin{figure}
\centering
\noindent
\includegraphics[width=10cm]{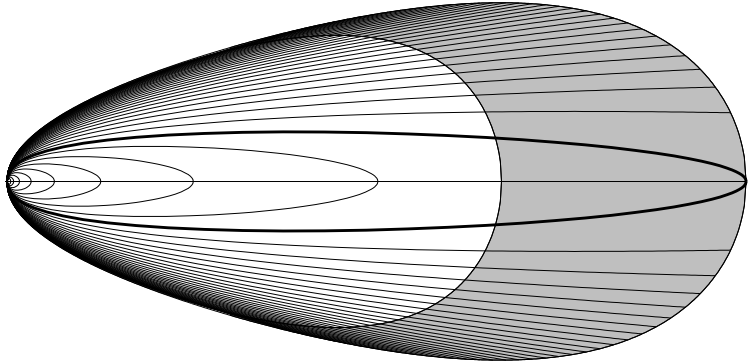}
\caption{\label{Fig5}
  Equal observed synchrotron frequency contour lines, at a given
  observed time, for the BM solution. Each line represents a different
  value of $\phi$. The bold contour line is $\phi=1$. The lines are
  separated by a constant logarithmic interval in $\phi$: $\Delta
  Log_{10}(\phi)=0.05$ (the inner contour lines, which cross the LOS,
  represent $\phi<1$). The shaded region represents a finite shell of
  thickness $\Delta = R/4\gamma^2$ in the observer frame. }
\end{figure}

These results can be best understood by looking at the ``egg'' that
constitutes the locus of points from which photons reach the detector
simultaneously, and mapping upon it the typical emitted frequency in
the observer frame. Figure \ref{Fig5} depicts the lines of equal
$\nu_{syn}$. From the electron distribution we obtain that $\langle
\gamma_e \rangle =\gamma_{min}(p-1)/(p-2) $ and therefore:
$\nu_{min}=(p-2)^2/(p-1)^2\nu_{syn}$. For this reason, the frequency
contour lines that are depicted in Figure \ref{Fig5} also represent
equal $\nu_{min}$ lines.

There are two opposing factors that determine the shape of the
observed synchrotron frequency contour lines, that are depicted in
Figure \ref{Fig5}. The shift from the local frame to the observer frame,
$\nu'=\nu\gamma(1-\beta\nu)$, causes the frequency to decrease as one
moves away from the LOS.  However at earlier emission times, and at
locations closer to the shock front, the typical synchrotron
frequencies in the local frame are higher, and so is the Lorentz
factor of the matter. This tends to increase the observed frequencies
of photons that were emitted earlier (i.e. from the back of the
``egg'') and closer to the shock (i.e.  closer to the surface of the
``egg'').

The result of these two opposing effects, for the BM solution, is that
as one goes backwards along the LOS, to earlier emission times, the
observed synchrotron frequency for a constant observed time is almost
independent of $\chi$: $\nu_{syn}(LOS)\propto \chi^{-1/24}$. This
explains the result for the flux arriving from the LOS (the dashed
curve in Figure \ref{Fig8}), namely that the peak flux is obtained at
a frequency just slightly lower than $\nu_{T}$: $\phi_{peak} \cong
0.98$, and the light curve is only slightly rounded near the peak. The
fact that $\nu_{min}<\nu_{T}$ along the LOS accounts for the fact that
the peak flux is obtained at $\phi_{peak}<1$. The fact that the
decrease in $\nu_{min}$ as one goes back along the LOS is very
moderate, means that in order for the emitted radiation to be
concentrated around a frequency substantially lower than $\nu_{T}$,
one has to go very deep in the radial profile along the LOS to $r \ll
R_l$. This implies that one gets far from the shock front (to $\chi
\gg 1$), and therefore the contribution obtained to the total flux is
small. For this reason $\phi_{peak}$ is very close to 1.

The observed flux from an infinitely thin shell peaks at $\phi_{peak}
\cong 1.92$, and the light curve (or spectra) is quite rounded and
flat near the peak (see the dotted line in Figure \ref{Fig8}). This
result can be understood by following the surface of the ``egg'' (see
Figure \ref{Fig5}).  $\nu_{min}$ increases as one goes to the back of
the ``egg'' (i.e. to earlier emission times) along it's surface, and
it does so much faster than it decreases when one goes back along the
LOS. Therefore one gets a substantial contribution to the flux at
$\phi>1$, before one gets too far back in the shell, where the total
contribution to the flux drops considerably. This explains why the
peak flux for an infinitely thin shell is obtained at a frequency
significantly higher than $\nu_{T}$, whereas for the LOS it is
obtained at a frequency only slightly lower than $\nu_{T}$.

\begin{figure}
\centering
\noindent
\includegraphics[width=13cm]{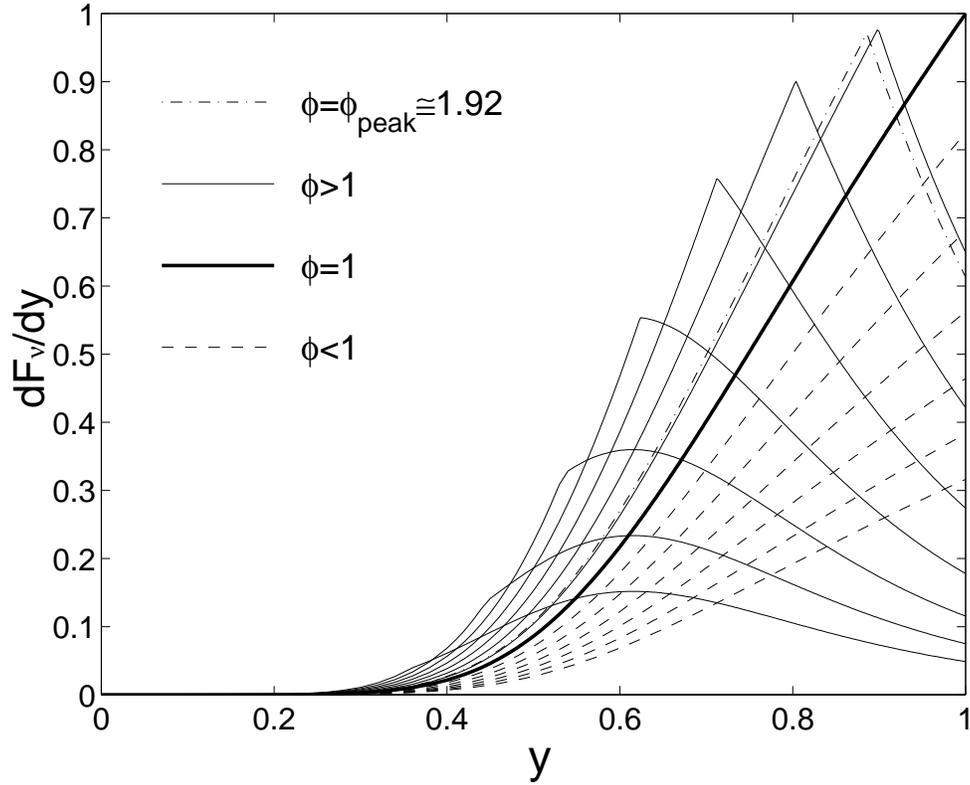}
\caption{\label{Fig9}
  The contribution to the observed flux as a function of $y\equiv
  R/R_l$, for an infinitely thin shell. The different curves stand for
  $Log_{10}(\phi)=-1.5,-1.25,...,1.75$ and $\phi = \phi_{peak} \approx
  1.92$. For $\phi<1$ the contribution peaks at the LOS ($y=1$), and
  for $\phi>1$ the contribution peaks at a lower value of $y$ as
  $\phi$ increases.}
\end{figure}

\begin{figure}
\centering
\noindent
\includegraphics[width=8cm]{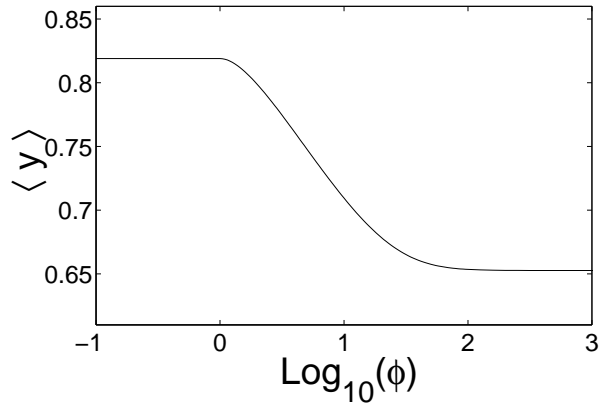}
\caption{\label{Fig10}
  The average normalized radius $\langle y \rangle$ contributing to
  the observed flux at a given $\phi$, for an infinitely thin shell.
  $\langle y \rangle \approx 0.82$ for $\phi<1$, and it decreases as
  $\phi$ increases above $\phi=1$, until it reaches the limiting value
  of $\langle y \rangle \approx 0.65$ for large $\phi$, at around
  $\phi \sim 100$.}
\end{figure}

The contribution to the flux from an infinitely thin shell, as a
function of the radius of the shell, for different values of $\phi$ is
shown in Figure \ref{Fig9}. The different curves can be looked upon as
representing either different observed frequencies at the same
observed time, or the same observed frequency at different observed
times. One can see that for $\phi<1$ the contribution to the flux
peaks at $y=1$ (the LOS), while for $\phi>1$ it peaks at smaller
values of $y$. One can obtain the average radius contributing to a
given $\phi$ in the following way:
\begin{equation}
\label{mean y}
\langle y \rangle={\int^1_0dy{dF(\phi) \over dy}y \over \int^1_0dy{dF(\phi)
\over dy}} \ .
\end{equation}
$\langle y \rangle (\phi)$ is shown in Figure \ref{Fig10}. For
$\phi<1$ one obtains that $\langle y \rangle$ is constant: $\langle y
\rangle \cong 0.82$, while for $\phi>1$ there is a decrease in
$\langle y \rangle$ until for $\phi \gg 1$ it reaches the limiting
value: $\langle y \rangle \cong 0.65$. The same infinitely thin shell
approximation was considered by Panaitescu \& M\'esz\'aros (1998) and
by Sari (1998). The values quoted by Panaitescu \& M\'esz\'aros for
$\langle y \rangle$ in the two limits of high and low frequencies are
probably erroneous, and they are not consistent with the value they
quote for the bolometric $\langle y \rangle$ (this inconsistency
occurs only for the case we consider in this paper, of an adiabatic
evolution of the shell, and neglecting electron cooling).  The values
obtained by Sari are slightly different than ours, since he considered
emission from a shell consisting of a given element of matter.

\section{Alternative Magnetic Field models}
\label{difB}
Although the hydrodynamic profile is given by the BM solution, the
structure and profile of the magnetic field are less clear. So far we
have assumed that everywhere the magnetic field energy is a fixed
fraction of the internal energy:
\begin{equation}
\label{equipartition}
e'_B \equiv {B^2 \over 8 \pi}=\epsilon_B e' \ .
\end{equation}
Since not much is known about the origin or spacial dependence of the
magnetic field, we consider now two alternative models for the
magnetic field. This can serve, in a way, as a test for the generality
of our results, and it will expose any hidden fine tuning caused due
to our previous magnetic field model, if one existed. We assume that
each matter element acquires a magnetic field according to equation
\ref{equipartition} when it passes the shock. The two alternative
models are obtained by assuming that the magnetic field is either
radial or tangential, and evolves according to the ``frozen field''
approximation.

Consider a small matter element, which passes the shock at a time
$t_0$ in the observer frame. Just after it passes the shock it
possesses a magnetic field $B_0$ (in the local frame) given by
equation \ref{equipartition}. We consider a cubic volume $V_0=L^3$ (in
the local frame) with one face perpendicular to the radial direction.
According to the BM solution, at a later time $t$ this matter element
will be at $\chi=(R(t)/R(t_0))^4$, and it will occupy a box of a size
$L_{rad}=\chi^{9/8}L$ in the radial direction, and a size
$L_{\perp}=\chi^{1/4}L$ in the two tangential directions ($\hat \phi$
and $\hat \theta$). One also obtains that $B_0=B_f \chi^{3/8}$, where
$B_f$ is the magnetic field at the front of the radial profile, just
behind the shock, at the time $t$.  We consider two possibilities for
the direction of the magnetic field at $t_0$: a radial and a
tangential magnetic field, $B_{rad}$ and $B_{\perp}$. Our previous
``equipartition'' model will be denoted simply by $B$. In both new
cases the ``frozen field'' approximation implies that the magnetic
field will remain in the same direction, while it's magnitude changes
in the following way:
\begin{equation}
\label{B1} 
B_{rad}=B_f \chi^{-1/8} \quad , \quad B=B_f \chi^{-17/24} \quad ,
 \quad B_{\perp}=B_f \chi^{-1} \ .
\end{equation}

\begin{figure}
\centering
\noindent
\includegraphics[width=11cm]{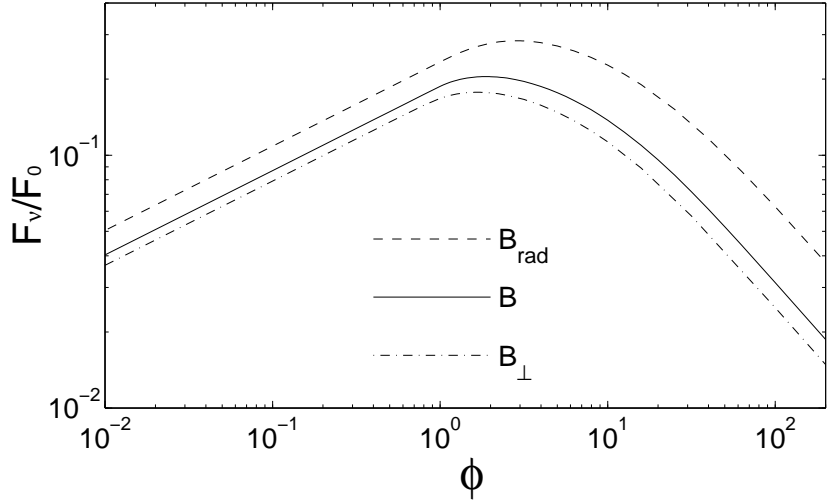}
\caption{\label{Fig11}
  The observed flux for three different magnetic field models.
  $B_{rad}=B=B_{\perp}$ on the shock front, while behind the shock
  front $B_{rad}>B>B_{\perp}$.}
\end{figure}

\newcommand{\rb}[1]{\raisebox{1.5ex}[0pt]{#1}}
\begin{table}
\begin{center}{\bf TABLE {\large 1}\\ \vspace{0.5cm}
\bf Features of the Light Curve and Spectra 
for\\ \vspace{0.3cm} Different Models for the Magnetic Field.}\vspace{0.5cm}
\begin{tabular}{|c||c|c|c|c|c|c||c|c|}\hline 
Magnetic & & & & & & & & \\ 
Field Model & \rb{\large $\phi_{peak}$} & 
\rb{\large{${F_{\nu,max}\over F_0}$}} & \rb{\large{$\phi_-$}} &
\rb{\large{$\phi_+$}} & \rb{\large{$\phi_+ \over \phi_-$}} &
\rb{\large{$T_+ \over T_-$}} &
\rb{\large{$F_{extr} \over F_0$}} & \rb{\large $\phi_{extr}$}  \\ \hline \hline
${B_{\perp}}$ & 1.67 & 0.178 & 0.157 & 15.9 & 101 & 21.7 & 0.273 & 4.12 
\\ \hline 
${B}$ &1.88 & 0.205 & 0.187 & 18 & 99 & 21.4 & 0.313 & 4.66 \\ \hline
${B_{rad}}$ & 2.86 & 0.284 & 0.241 & 29 & 120 & 24.3 & 0.454 & 7.30 \\ \hline
\end{tabular}\\
\end{center}
\caption{$\phi_{peak}$, $F_{\nu,max}$, $F_0$, $\alpha$, $\phi_-$,
  $\phi_+$, $T_-$ and $T_+$ are all defined in the text. 
  The last two columns refer to the point where the 
  extrapolations of the power laws at high and low $\phi$ meet.}
\end{table}

Using the general formula (equation \ref{FAF}), we obtain now the
observed flux for $B_{rad}$ and $B_{\perp}$ (see Figure \ref{Fig11}).
Some of the features are summarized in Table 1. Since the emission
from the shock front ($\chi=1$) is identical in all three cases, the
results obtained for an infinitely thin shell in the previous section,
including the dotted curve in Figure \ref{Fig8}, are still valid here.
This implies that the difference between the various curves in Figure
\ref{Fig11} arises from the radial integration. Our previous model for
the magnetic field led to an almost negligible effect of the radial
profile (see Figure \ref{Fig8}). Now we find that generally, the
effect of the radial integration is comparable to that of the angular
integration, as the flux for $B_{rad}$ and $B_{\perp}$ are
substantially different than for an infinitely thin shell.

From equation \ref{B1}, we can see that $B_{rad}$ implies a larger
magnetic field than $B$, and therefore a larger total emission and
higher emission frequencies, while $B_{\perp}$ implies a smaller
magnetic field than $B$, and therefore a smaller total emission and
lower emission frequencies.  This is consistent with the results in
Table 1, namely: $\phi_{peak,rad}>\phi_{peak}>\phi_{peak,\perp}$ and
$F_{\nu,max,rad}>F_{\nu,max}>F_{\nu,max,\perp}$.  An important feature that
appears in all the magnetic field models we considered, is the
flatness of the peak: $\phi_+/\phi_-\sim 100-120$ and $T_+/T_- \sim
21-24$ (see Table 1).

\section{A Uniform Shell Approximation}
\label{Eli}
In this section we consider a locally monochromatic emission. We
consider a case where all the electrons at a certain point have the
same Lorenz factor, which we chose to be the average Lorenz factor of
the electrons at that point, given by equation \ref{P el}. We take the
angle $\alpha$ between the velocity of the electrons and the direction
of the magnetic field in the local frame, to be the average value
obtained for an isotropic velocity distribution: $\sin\alpha=\pi/4$.
Under these assumptions, the emission from this point is at a single
frequency, given by equation \ref{typical syn freq}. The results are
still expressed by the similarity variable $\phi \equiv \nu/\nu_T$,
where in this section $\nu_T$ can be obtained from equation \ref{nuT}
by dropping the term involving $p$. The emitted power per unit volume
per unit frequency in the local frame is given by:
\begin{equation}
\label{delta emission function}
P'_{\nu'}=n' P'_{e,av} \delta (\nu'-\nu'_{syn}) \ ,
\end{equation}
where $P'_{e,av}$ is given by equation \ref{P el}.

\begin{figure}
\centering
\noindent
\includegraphics[width=9cm]{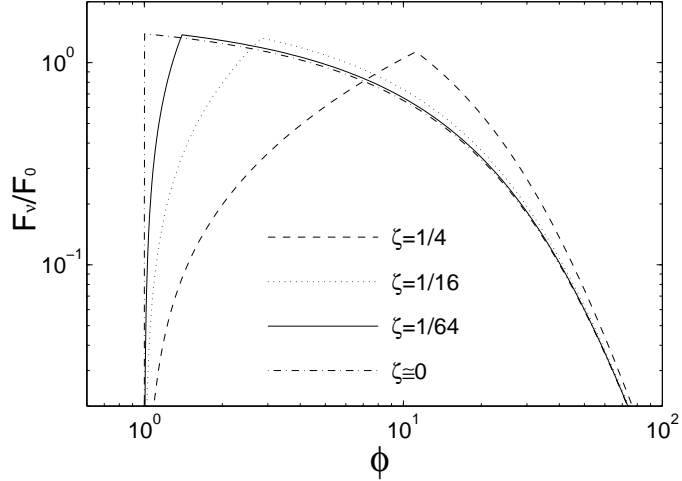}
\caption{\label{Fig7}
  The observed flux from a uniform shell, for different values of the
  shell thickness in the observer frame: $\Delta \equiv \zeta
  R/\gamma^2$.  The value of $\phi$ at the peak is a function of the
  shell thickness (see equation \ref{peak flux}).}
\end{figure}

We now turn to calculate the observed flux due to this emission within
a thin shell of matter, of finite thickness $\Delta$ in the observer
frame. This case was calculated numerically by Waxman (1997c). He
took: $\Delta=\zeta R/\gamma^2$, which corresponds in our notation to:
$\chi_{max}(\zeta)=1+16\zeta$. We use equation \ref{FAF}, and
integrate only up to $\chi_{max}(\zeta)$. Substituting $g(\chi)=1$ in
equation \ref{FAF}, We obtained an analytic solution, and present it
in Figure \ref{Fig7}. We compare this both with a locally
monochromatic emission from a BM profile (see Figure \ref{Fig4}), and
with synchrotron emission from a BM profile, obtained in section
\ref{ni3}.

\begin{figure}
\centering
\noindent
\includegraphics[width=12cm]{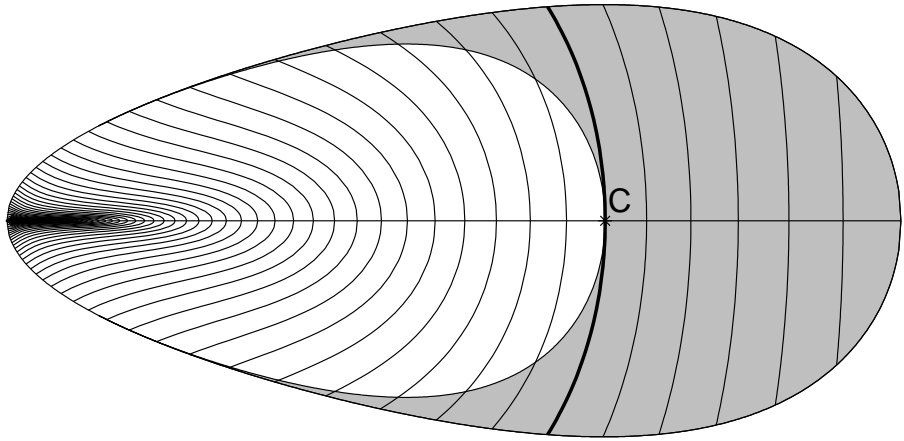}
\caption{\label{Fig6}
  Equal observed synchrotron frequency contour lines, at a given
  observed time, for a uniform shell. The lines from right to left,
  represent: $\phi=5^{1/4},5^{2/4},5^{3/4},...$. The typical
  synchrotron frequency is larger than $\nu_T$ everywhere (everywhere
  $\phi>1$), and $\phi$ increases as one goes to the back (i.e. to the
  left) of the ``egg'' along the LOS, to smaller radii.  The shaded
  region represents a finite shell of thickness $\Delta = R/4\gamma^2$
  in the observer frame. The bold frequency contour line that touches
  the back of the shell on the LOS at point C, represents the
  frequency for which the flux from a uniform shell, considered in
  this section, peaks.}
\end{figure}

\begin{figure}
\centering
\noindent
\includegraphics[width=8cm]{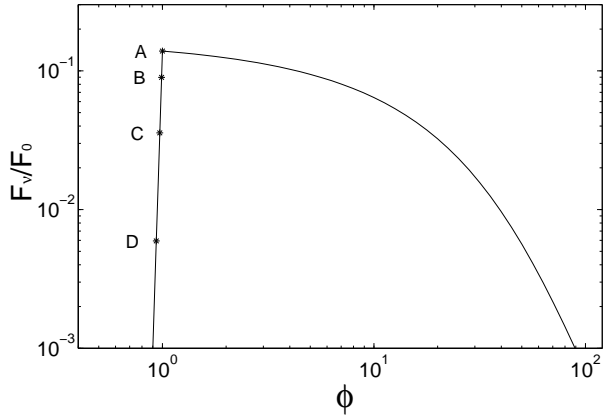}
\caption{\label{Fig4}
  The observed flux from a locally monochromatic emission at the
  typical synchrotron frequency, using the BM solution. Since $\phi
  \propto \nu T^{3/2}$, this figure represents (with the correct
  scaling) both the light curve at a given frequency, and the spectrum
  at a given observed time. The peak flux is obtained at $\phi=1$, and
  the flux drops sharply for smaller $\phi$. If we consider only a
  slice of a BM profile, this introduces a cutoff at low $\phi$. This
  cutoff occurs at points A through D, for $\zeta \to 0$,
  $\zeta=1/64$, $\zeta=1/16$ and $\zeta=1/4$, respectively, where the
  shell thickness in the observer frame is $\Delta\equiv\zeta
  R/\gamma^2$.}
\end{figure}

The flux obtained from a finite shell with a ``cut'' BM profile, is
very similar to the flux from a full BM profile, shown in Figure
\ref{Fig4}, except that there is a cutoff at
$\phi<(1-16\zeta)^{-1/24}$, due to the finiteness of the shell. The
place of this cutoff is shown in Figure \ref{Fig4}, for a few values
of $\zeta$, including $\zeta \to 0$. Figure \ref{Fig7} depicts the
observed flux from a uniform shell, for the same values of $\zeta$,
and is considerably different. This large difference can be understood
by looking at Figures \ref{Fig5} and \ref{Fig6}, which depict the
contour lines of equal observed emission frequency (which will be
referred to as frequency contour lines) for the BM solution and for a
uniform shell, respectively.

We now turn to Figure \ref{Fig4}, which depicts the flux from a BM
profile with a locally monochromatic emission. Normalizing according
to the maximal flux, the flux from the LOS and the flux from an
infinitely thin shell, practically coincide with the flux calculated
from the general formula, for $\phi<1$ and $\phi>1$, respectively, and
therefore are not shown separately. The flux peaks at $\phi_{peak}=1$
and drops sharply for $\phi<1$, while exhibiting a more moderate
decrease for $\phi>1$. The frequency contour lines for the BM solution
are nearly parallel to the LOS and their length within the thin shell
(or up to a constant value of $\chi$) drops quickly to zero for
$\phi<1$ (see Figure \ref{Fig5}). This explains the quick drop (and
eventual cutoff, for a finite shell) of the flux for $\phi<1$, and the
fact that the peak flux is located at $\phi=1$. If one follows the
shock front (i.e. the surface of the ``egg''), one finds that the
observed frequency increases as one moves from the LOS to the back of
the ``egg'' (i.e.  to smaller radii). This explains why we find a
cut-off in the flux arriving from an infinitely thin shell, at
frequencies smaller than $\nu_{T}$ ($\phi<1$). Since there is a sharp
drop in the contribution to the flux as one goes back in the radial
profile (away from the shock front), the flux at $\phi>1$ where there
is a contribution from the front of the radial profile, is similar to
the one obtained for an infinitely thin shell.  For $\phi<1$,on the
other hand, there is no contribution from the very front of the radial
profile (near the shock front), and therefore the contribution from
back the radial profile becomes apparent, and we obtain a result
similar to the one obtained for the LOS.

The observed flux for a uniform shell is shown in figure \ref{Fig7}.
The shape of the light curve and the spectra, the shape and sharpness
of the peak and the value of $\phi_{peak}$, all change considerably
for different values of the shell thickness $\zeta$. The big
difference compared to the results for a BM profile, can be understood
by looking at Figure \ref{Fig6}, which depicts the frequency contour
lines for a uniform shell. The frequency contour lines are nearly
perpendicular to the LOS and higher frequencies come from the back of
the ``egg'' (i.e.  from smaller radii). For this reason the flux
vanishes for $\phi<1$ (see Figure \ref{Fig7}).  For every value of
$\zeta$ there exists a critical frequency $\phi_c(\zeta)$ for which
the frequency contour line touches the back of the shell, exactly on
the LOS. This is demonstrated by the bold frequency contour line in
Figure \ref{Fig6}, which represents $\phi_c$ for $\zeta=1/4$, and
touches the back of the shell exactly on the LOS at point C. For
$\phi<\phi_c(\zeta)$ there is a sharp drop in the length of the
frequency contour lines within the thin shell, that causes a sharp
drop in the flux. For $\phi>\phi_c(\zeta)$ there is a moderate change
in the length of the frequency contour lines within the thin shell,
and the flux drops (though more moderately) for $\phi>\phi_c(\zeta)$
as well. This causes the peak flux to be exactly at $\phi_c(\zeta)$.
From our analytical solution for a uniform thin shell we obtained a
simple analytical expression for the time (or frequency) of the peak
flux as a function of $\zeta$:
\begin{equation}
\label{peak flux}
\phi_{peak}(\zeta)=\phi_c(\zeta)=(1+16\zeta)^{3/2} \ ,
\end{equation}

The resulting spectra and light curve for a uniform shell are
qualitatively different from those obtained when the hydrodynamical
radial profile is considered. For a uniform shell, the rather
arbitrary choice of $\zeta$ determines the time of the peak flux for a
given observed frequency, and the peak is substantially sharper, with
a smaller width at half maximum. The difference is even bigger if the
results for a uniform shell are compared to those obtained for a
realistic emission and a realistic hydrodynamic solution (the solid
curve in Figure \ref{Fig8}).

\begin{figure}
\centering
\noindent
\includegraphics[width=13cm]{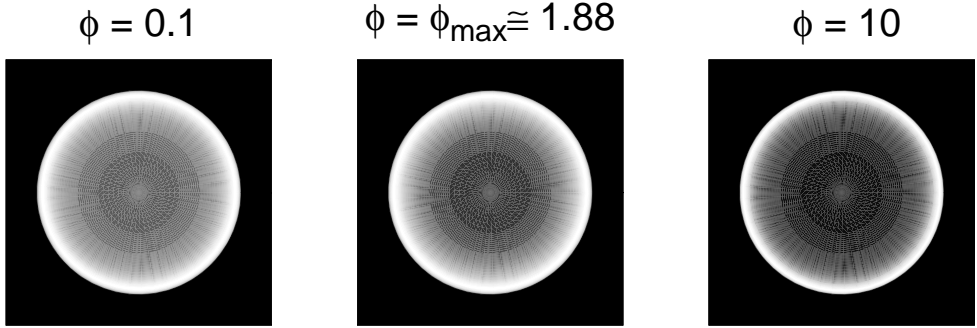}
\caption{\label{Fig12}
  The observed image of a GRB afterglow, for different values of
  $\phi$. The different images can represent either the image at
  different frequencies at the same observed time, or the image at the
  same frequency at different observed times. In the latter case, one
  should also remember that the size of the observed image increases
  with the observed time as $R_{\perp,max} \propto T^{5/8}$. A larger
  value of $\phi$ represents a higher observed frequency or a later
  observed time. The image is brighter near the edge and dimmer near
  the center, and the contrast is higher at large values of $\phi$.}
\end{figure}

\section{The Observed Image}
We turn now to calculate the observed image, as seen by a distant
observer. We consider the BM hydrodynamic solution. Substituting
$g(\chi)=\chi^{-1}$ in equation \ref{FAF} we obtain:
\begin{equation}
\label{dF} 
dF_{\nu}={8 R_l^3 \over D^2}{P'  y^{10} \chi \over (1+7\chi y^4)^2 }dyd\chi \ .
\end{equation}
We calculate the surface brightness (energy per unit time per unit
frequency per unit area perpendicular to the LOS) at a given arrival
time $T$. The distance of a point from the LOS is given by:
\begin{equation}
\label{Rp} 
R_{\perp} \equiv r\sin \theta \cong R_l y \sqrt {1-\mu^2}
\cong {\sqrt2 R_l \over 4\gamma_l} \sqrt {y-\chi y^5} \ .
\end{equation}
The maximal value of $R_{\perp}$ is obtained on the surface of the
``egg'' (where $\chi=1$), implying $y=5^{-1/4}$.  The observed image
at a given observed time $T$ is restricted to a disk of radius
$R_{\perp,max} \cong 4.14 \gamma_l c T$ around the LOS. Using equation
\ref{BM2}, we obtain an explicit expression for the BM solution:
\begin{equation}
\label{Rpmax} 
R_{\perp,max}= 3.91\times10^{16}E_{52}^{1/8}n_1^{-1/8}T_{days}^{5/8}\rm{cm}\ .
\end{equation}
We calculate the surface brightness within this disk, and find it
useful to work with the variable: $x \equiv R_{\perp}/R_{\perp,max}$.
The differential of the area on this disk is given by:
\begin{equation}
\label{dSp} 
dS_{\perp}=2\pi R_{\perp}dR_{\perp} \equiv 2 \pi R^2_{\perp,max} xdx \ .
\end{equation}
Using equation \ref{Rp}, we obtain from equation \ref{dF}:
\begin{equation}
\label{dFdSp} 
{dF_{\nu} \over dS_{\perp}}={4 \over \pi}\left({R_l \over D}\right)^2
{1\over c T}{P' y^5 \chi \over (1+7\chi y^4)^2}dy \ ,
\end{equation}
from which we obtain after integration, the surface brightness as a
function of $R_{\perp}$. In this expression, $\chi$ should be taken as
a function of $y$ for a given $x$, according to equation \ref{Rp}.
The limits of the integration over $y$ are determined
from the condition $\chi>1$.

\newcommand{\rbb}[1]{\raisebox{1.5ex}[0pt]{#1}}
\begin{table}
\begin{center} 
{\bf {TABLE \large 2}\\ \vspace{0.5cm} Features of the Observed Image.\\}
\vspace{0.5cm}
\begin{tabular}{|c||c|c||c|c||c|c|}\hline
Magnetic &\multicolumn{2}{c||}{\ } & \multicolumn{2}{c||}{\ }
& \multicolumn{2}{c|}{\ } \\ 
Field   &\multicolumn{2}{c||}{\rbb{${\rm SB}(x=0)/{\rm SB}_{max}$}} &
\multicolumn{2}{c||}{\rbb{$\phi < 1$}} &
\multicolumn{2}{c|}{\rbb{$\phi \gg 1$}} \\ \cline{2-7}
Model & $\phi<1$ & $\phi \gg 1$ & 
$x_{peak}$ & $\Delta x$ & $x_{peak}$ & $\Delta x$ \\ \hline \hline
${B_{\perp}}$ & 0.32 & 0.034 & 0.95 & 0.263
 & 0.96 & 0.151 \\ \hline 
${B}$ & 0.34 & 0.039 & 0.94 & 0.296
& 0.95 & 0.178 \\ \hline
${B_{rad}}$ & 0.39 & 0.065 & 0.93 & 0.403
& 0.90 & 0.290 \\ \hline
\end{tabular}\\
\end{center}
\caption{. We define $\Delta x \equiv x_+ - x_-$, where $x_+>x_-$ are
the values of $x\equiv R_{\perp}/R_{\perp,max}$ at which
the surface brightness (SB) drops to half of it's maximal
value. $\Delta x$ is an estimate for the width of the bright ring 
that appears on the outer edge of the image. $x_{peak}$ is the value
of $x$ for which the maximal SB is obtained, and it's
values indicate that the SB peaks near the outer edge of the image.
It is evident that the contrast between the center and the edge of the 
image is considerably larger for high frequencies (or late times) than 
for low frequencies (or early times).}
\end{table}

\begin{figure}
\centering
\noindent
\includegraphics[width=13cm]{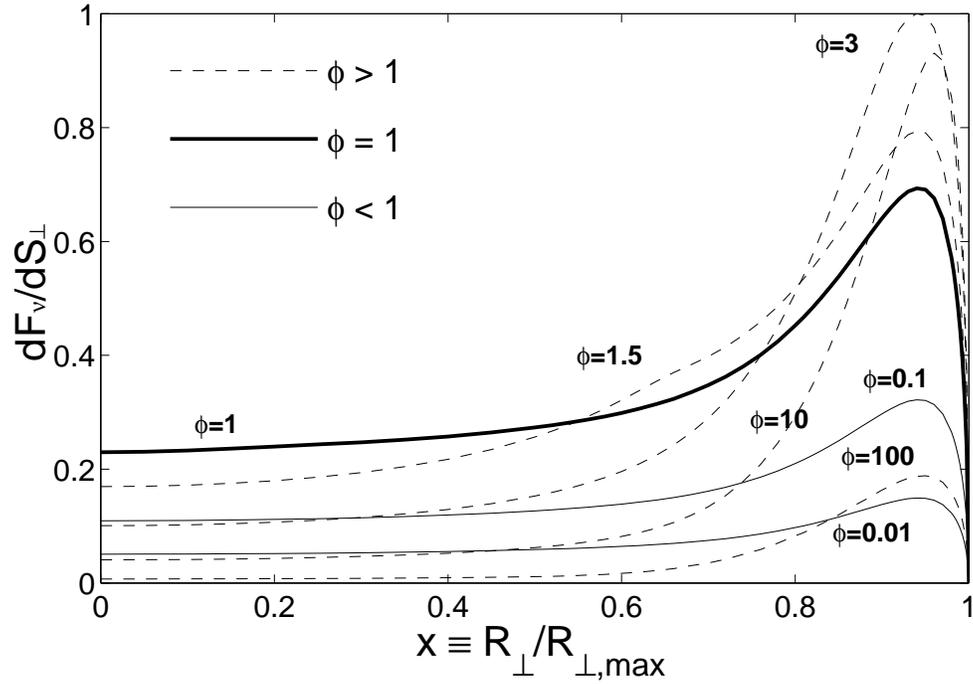}
\caption{\label{Fig13}
  The observed surface brightness as a function of $x \equiv
  R_{\perp}/R_{\perp,max}$, for different values of $\phi$.  The
  surface brightness peaks near the outer edge of the image, drops
  sharply towards the edge ($x=1$) and decreases more gradually
  towards the center. The normalization preserves the correct ratio
  between the absolute surface brightness, for the different values of
  $\phi$, which decreases both for low and for high values of $\phi$.}
\end{figure}

We calculated the surface brightness for several values of $\phi$, and
as before, the different images can be viewed as either the observed
images at a given observed time at different observed frequencies, or
as the observed images at a given frequency at different observed
times. The observed images are quite similar for the different
magnetic field models we considered, and therefore we present detailed
results only for $B$ (figures \ref{Fig12} and \ref{Fig13}), and
summarize the features of the images obtained for $B_{rad}$, $B$ and
$B_{\perp}$ in Table 2.

The observed image is bright near the edge and dimmer at the center.
At high values of $\phi$ (high frequencies or late observed times) the
surface brightness at the center is only a few percent of the maximal
surface brightness, which is obtained near the edge ($x \sim
0.93-0.95$), and a bright ring is clearly seen with a sharp edge on
the outer side and a more gradual decrease in the surface brightness
towards the center.  At $\phi<1$ (low frequencies or early observed
times) the surface brightness at the center is around $0.32-0.39$ of
the maximal surface brightness, and though the image is brighter near
the edge, the center should be visible as well (see the left image in
Figure \ref{Fig12}).

The transition in the distribution of the relative surface brightness
along the observed image, between the limiting cases of small and
large $\phi$, occurs over one order of magnitude in $\phi$:
$1<\phi<10$.  The relative surface brightness hardly changes beyond
this region (though the absolute surface brightness drops both for low
and for high values of $\phi$, due to the drop in the observed flux).

The fact that at a given observed time, the relative surface
brightness at the center of the image is smaller for high frequencies
than it is for low frequencies, may be intuitively understood: for
high frequencies the emittance decays in time, and is therefore lower
at large radii, which correspond to small values of $x \propto
R_{\perp}$, which are located at the center of the image.

\section{Discussion}
We have derived a formula for the observed flux from a
relativistically moving system. Using this formula, we considered the
emission from the whole volume behind a spherical relativistic blast
wave, expanding into a cold uniform medium. We have considered an
adiabatic evolution and used the self similar hydrodynamical solution
of Blandford \& McKee (1976) to describe the matter behind the
shock. 

We have considered synchrotron emission from a power law distribution
of electrons, using a realistic local emissivity, which depends on the
local hydrodynamic parameters. We have obtained expressions for the
frequency at which the observed flux peaks at a given observed time,
and for the value of the peak flux (equations \ref{ni peak} and
\ref{Fmax}). The value we obtained for the peak frequency at a given
observed time is a factor of $\sim 1.8$ smaller than the value
obtained by Sari, Piran \& Narayan (1998), a factor of $\sim 2.5$ (for
$p=2.5$) smaller than the value obtained by Wijers \& Galama (1998),
and a factor of $\sim 35$ (for $p=2.5$) smaller than the value
obtained by Waxman (1997b). The large difference from the result
obtained by Waxman is partly due to the fact that he took the peak
frequency to be at the synchrotron frequency of an average electron
$\nu_{syn}(\langle \gamma_e \rangle)$, while the peak frequency is
actually obtained much closer to $\nu_{syn}(\gamma_{min})$, i.e.  the
synchrotron frequency of an electron with the minimal Lorentz factor.
The value we obtained for the peak flux is a factor of $\sim 1.9$ (for
$p=2.5$) smaller than the value obtained by Wijers \& Galama (1998), a
factor of $\sim 5.1$ larger than the value obtained by Waxman (1997b),
and a factor of $\sim 6.5$ smaller than the value obtained by Sari,
Piran \& Narayan (1998).

We have shown that both the light curve and the spectra are flat near
the peak. The flux at a given observed time drops to half it's maximal
value at around one order of magnitude from the peak frequency, on
either side. The flux at a given observed frequency drops to half it's
maximal value at about a factor of $\sim 5$ before and after the time
of the peak flux. This result was obtained for all the magnetic field
models we considered, and it therefore seems to be of quite a general
nature. Our calculations were made using an approximation for the
local emissivity that is obtained from an exact synchrotron emission.
We expect that without this approximation the values of the peak flux
and the peak frequency should be somewhat lower, and the light curve
and the spectra should be even more rounded and flat near the peak.
We estimated that these effects should be small.

In contrast to the flatness of the peak, discussed above, GRB970508
displayed a sharp rise in the optical flux, immediately followed by a
power law decay (Sokolov et. al. 1997). Sari, Piran \& Narayan (1998)
considered both radiative and adiabatic evolution of the blast wave,
and found that the steepest rise in the flux occurs before the peak,
for an adiabatic evolution and slow cooling of the electrons, which is
the case discussed in this paper.  This steepest rise is $T^{1/2}$,
and as we have shown in this paper, the rise in the flux decreases as
the peak is approached, and the peak itself is quite flat. We obtained
$T_+/T_- \sim 21-24$, which indicates a flat peak, while GRB970508
displayed $T_+/T_- < 3$. This rules out the interpretation of the
optical peak of GRB970508 as the peak of the light curve, predicted by
the existing fireball models. It therefore appears that another
explanation is needed.

We have considered a locally monochromatic emission, at the typical
synchrotron frequency, obtained from the relevant hydrodynamic
parameters. We have shown that when this emission is applied to the BM
solution, the observed flux peaks at the observed emission frequency
of the matter just behind the shock, on the LOS ($\phi_{peak}=1$).
There is a sharp drop in the flux below this frequency, and a gradual
decrease above it.  When this emission is applied to a uniform shell
(Waxman 1997b) the results change drastically. In particular the
location of the peak flux depends critically on the width of the shell
$\zeta$ (see equation \ref{peak flux}), which is chosen quite
arbitrarily.

The image of a GRB afterglow looks like a ring, even when emission is
considered from the whole volume behind the shock front. Similar
results were obtained for more simplified models (Waxman 1997b, Sari
1998, Panaitescu \& M\'esz\'aros 1998). The image is bright near the
edge and dimmer at the center. The contrast in the surface brightness
between the center and the edge of the image is larger for high
frequencies (optical and x-ray) and the ring is narrower, while for
low frequencies (as long as self absorption is not significant) the
contrast is smaller and the ring is wider. Radio wave-bands can be
considered as ``low frequencies'' for the first few months. The best
available resolution is obtained in radio frequencies, and the
afterglow of a future nearby GRB ($z\sim 0.2$) might be resolved. This
theory predicts that in early times, when $\nu_{radio}<\nu_{peak}$ the
image should appear as a relatively wide ring with a relatively small
contrast, while for later times where $\nu_{radio}>\nu_{peak}$ (as
long as the relativistic regime is not exceeded) the image should
appear as a narrow ring and posses a large contrast.

The observed image has been calculated, considering emission only from
the surface of the shock front (Sari 1998). This yielded a
surface brightness diverging at $R_{\perp,max}$. This divergence is an
artifact of the assumption that the radiation is emitted from a two
dimensional surface. Other features of the image, such as the
difference between high and low frequencies, are quite similar in the
more simplified analysis.

\vspace{1cm}
This research was supported by NASA Grant NAG5-3516, and a US-Israel
Grant 95-328. Re'em Sari thanks The Clore Foundation for support.

\end{document}